\documentclass[journal]{IEEEtran}
\usepackage{graphicx}
\usepackage{amsmath, amssymb}
\ifCLASSINFOpdf
\usepackage{float, multirow}
\else
\usepackage{cite}
\fi
\usepackage{xcolor}
\usepackage{cite}
\usepackage{hyperref}
\hypersetup{%
  citebordercolor=blue
}
\usepackage{booktabs}
\usepackage{balance}

\begin{document}

\title{Noise-Domain Non-Orthogonal Multiple Access for Three Users}

\author{
  Erkin Yapici,~\IEEEmembership{Graduate Student Member,~IEEE,} Yusuf Islam Tek,~\IEEEmembership{Graduate Student Member,~IEEE,} \\  Ali Emre Pusane,~\IEEEmembership{Senior Member,~IEEE}
  and Ertugrul Basar,~\IEEEmembership{Fellow,~IEEE}%
\thanks{E. Yapici is with the Department of Electrical and Electronics Engineering, Bogazici University, Istanbul,Türkiye. Email: erkin.yapici@std.bogazici.edu.tr}%
\thanks{Y. I. Tek is with the Department of Electrical and Electronics Engineering, Koc University, Sariyer, Istanbul, Türkiye, and the R\&D Department, Turk Telekom, Ankara, Türkiye. Email: ytek21@ku.edu.tr and\\ yusufislam.tek@turktelekom.com.tr}%

\thanks{Ali Emre Pusane is with the Department of Electrical and Electronics Engineering, Bogazici University, Istanbul, Turkiye. Email: ali.pusane@bogazici.edu.tr}
\thanks{
E. Basar is with the Department of Electrical Engineering, Tampere University, 33720 Tampere, Finland, on leave from the Department of Electrical and Electronics Engineering, Koc University, Sariyer, Istanbul, Türkiye. Email: ertugrul.basar@tuni.fi and ebasar@ku.edu.tr }
\thanks{This work is supported by the Scientific and Technological Research Council of Türkiye (TÜBİTAK) through the 1515 Frontier R\&D Laboratories Support Program for the Türk Telekom 6G R\&D Lab (Project No. 5249902) and under Grant No. 124E146.}
}

\maketitle

\begin{abstract}
 In this study, we propose a novel three-user noise-domain non-orthogonal multiple access (ND-NOMA) scheme by introducing the correlation as a new dimension besides mean and variance quantities used in two-user ND-NOMA. The new three-user ND-NOMA scheme includes both uplink and downlink scenarios, with detectors designed to decode the information embedded in mean, variance, and correlation. Our theoretical analysis and simulation results under Rician fading channels show that the proposed system is capable of achieving promising bit error rate (BER) performance while preserving the low power and low complexity advantages of ND-NOMA. This new ND-NOMA design enables simultaneous communication among three users using different dimensions, paving the way for scalable multi-user communication in noise-domain systems and in the Internet-of-things (IoT) environments.
\end{abstract}

\begin{IEEEkeywords}
Non-orthogonal multiple access (NOMA), noise-domain modulation, ND-NOMA, Rician fading, low-complexity receivers, IoT networks, multi-user systems.
\end{IEEEkeywords}

\IEEEpeerreviewmaketitle

\section{Introduction}
Non-orthogonal multiple access (NOMA) has appeared as a key technology for beyond 5G, addressing the increasing demands for massive connectivity, low latency, and improved spectral efficiency. By enabling multiple users to simultaneously share the same time and frequency resources, NOMA succeeded in breaking the traditional constraints of orthogonal access and opened the way for better user deployment \cite{8357810,9691334, ahmed2024unveiling}. One of the most popular forms, power-domain NOMA (PD-NOMA), separates users by assigning different power levels to the users. At the receiver side, successive interference cancellation (SIC) is employed to iteratively decode and subtract stronger signals first, and then the weaker ones \cite{7676258}. Even though the PD-NOMA offers theoretical gains in spectral efficiency and user fairness, it suffers from practical limitations including SIC complexity and sensitivity to channel estimation errors.

Back in foundational research, to carry information, thermal noise modulation (TherMod) was introduced \cite{kish2005stealth, basar2023thermal, zayyani2025resistorhoppingkljnnoise}. On top of this foundational concept, subsequent studies proposed modulating information through the statistical parameters of the noise signal itself, such as the variance or mean \cite{10373568, anand2019wireless, yapici2025noisemodulationwirelessenergy, dos2025off, 11071299, da2026survey, 11242237, ma2025initial, 11357526}. These studies jointly demonstrate that noise modulation has the potential for simplifying receiver design, increasing robustness towards the channel imperfections, achieving covert and secure communications, enabling low complexity communication. Specifically, certain noise modulation forms  allow for non-coherent detection, where the receiver can extract information solely by estimating signal statistics such as variance, without the need of any knowledge of the instantaneous channel state. Furthermore, phase and frequency synchronization is not required in these schemes, which significantly reduces the receiver complexity. To add on, by distributing the noise samples corresponding to a single information bit across multiple time slots, noise modulation inherently enables time diversity, which enhances robustness in fading environments without requiring complex combining techniques or multiple antennas. Finally, noise-alike waveforms provide significant advantages in terms of secure and covert communications. These features make noise modulation suitable for low-data-rate scenarios, where minimizing receiver complexity and maximizing energy efficiency are critical. From a practical implementation perspective, the ideal white-noise assumption can be relaxed by considering colored noise with a controlled power spectral density, which may better represent realizable noise waveforms in practical RF systems. Moreover, practical transmitters and receivers inherently require appropriate filtering to limit the occupied bandwidth and shape the generated noise spectrum.

Building on this foundation and the fascinating property of the Gaussian distribution, where the sum of two independent Gaussian variables remains Gaussian \cite{6868214}, our prior work introduced a two-user noise-domain non-orthogonal multiple access (ND-NOMA) scheme, where user information was encoded through the mean and variance of artificially generated noise, offering a promising solution for emerging energy-constrained applications \cite{yapici2024noisedomainnonorthogonalmultipleaccess, butt2023ambientiotmissinglink}. Distinct from PD-NOMA schemes that rely heavily on SIC, ND-NOMA inherently avoids the need for such complex receiver operations by employing statistical noise parameters for information encoding. Moreover, as the scheme operates purely in the noise domain, it eliminates the requirement for precise frequency and phase synchronization, further reducing receiver complexity and making it highly attractive for low-data-rate applications. On the other hand also, ND-NOMA system has a simple detection structure thus it avoids typical error propagation problems which we observe in PD-NOMA systems. This foundational progression sets the stage for our current work, where we reshape the ND-NOMA paradigm to support a third user by leveraging correlation as a new statistical dimension, thereby enhancing the scalability and flexibility of noise-domain communication. It is worth noting that some earlier schemes considered correlation to carry information in secure and covert noise-driven communications, but not in the context of multiple accessing \cite{salbert1999secure}.

\begin{figure*}[!t]
    \centering
    \includegraphics[width=0.8\linewidth]{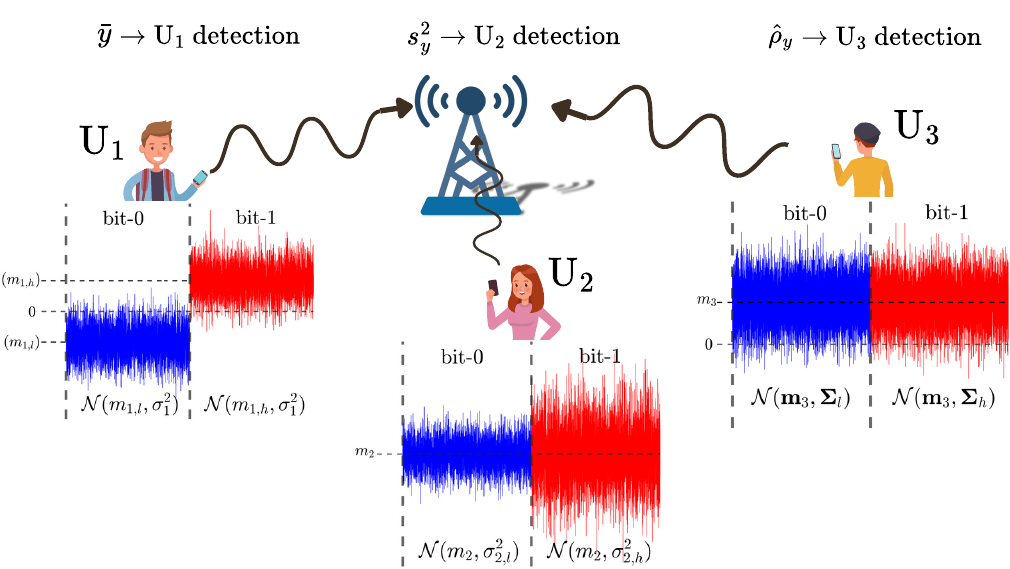}
    \caption{Uplink ND-NOMA scheme with three users using real Gaussian signals.}
    \label{fig:uplink}
\end{figure*}

In this work, we propose a novel ND-NOMA framework that enables simultaneous communication among three users by utilizing independent statistical properties of the Gaussian noise: mean, variance, and correlation. In this setup, the first user embeds information into the mean of the noise samples, the second user manipulates the variance, and the third user cleverly modulates the correlation between consecutive noise samples to convey data. We present system models for both uplink and downlink scenarios and develop dedicated detection mechanisms customized to each user’s encoding strategy. Simulation results and theoretical analysis are provided under Rician fading channels to evaluate bit error rate (BER) performance. The numerical results confirm that our proposed scheme retains the low BER and low-complexity characteristics over Rician fading channels, while significantly enhancing multi-user support. Hence, our proposed scheme specifically reduces the system complexity, which enables suitability for Internet-of-Things (IoT) environments. In order to support reliable multi-user communication under the power constraints of future IoT cities, the three-user ND-NOMA system uses a diverse detection architecture: mean-based detection for the first user, variance thresholding for the second user, and correlation detection for the third user. This layered detection structure of the three-user ND-NOMA significantly enhances robustness, maintaining reliability even under challenging channel conditions, while entirely eliminating the need for SIC thanks to its orthogonal modulation over mean, variance, and correlation.

The rest of the paper is organized as follows. Section II presents the system model and detection rules for the uplink scenario. Section III extends the analysis to the downlink case. Section IV provides simulation results and performance discussions. Finally, Section V concludes the paper and outlines potential directions for future research.

%\section{Downlink ND-NOMA: System Model and Performance Analysis}

\section{Uplink Three-User ND-NOMA: System Model and
Performance Analysis}
In this section, we introduce the three-user ND-NOMA scheme. In the proposed scheme, each user utilizes a distinct noise-domain property to encode data: mean, variance, and correlation. Building on the two-user ND-NOMA framework, we extend the model by introducing a third user whose information is embedded in the correlation between pairs of Gaussian noise samples. We also provide a general framework to evaluate the uplink BER performance for all users under Rician fading channel scenarios.

%\begin{figure}[!ht]
 %   \centering
  %  \includegraphics[width=\linewidth]{downlink.png}
   % \caption{Downlink scheme.}
    %\label{fig:downlink}
%\end{figure}

\subsection{System Model}
As illustrated in Fig.~\ref{fig:uplink}, the uplink transmission of the three-user ND-NOMA system encodes the information through the distinct statistical features of the Gaussian noise. To be specific, User~1 (U$_1$) modulates the mean, User~2 (U$_2$) modulates the variance, and User~3 (U$_3$) modulates the correlation between the Gaussian samples. In our new scheme, each user transmits its bit using $N$ real Gaussian noise samples per bit period.

For U$_1$, information bits are transmitted by altering the mean of Gaussian samples, while keeping the variance constant. That is, for bit-0 and bit-1, we have $s_1^n \sim \mathcal{N}(m_{1,l}, \sigma_1^2)$ and $s_1^n \sim \mathcal{N}(m_{1,h}, \sigma_1^2)$, with fixed variance $\sigma_1^2$ and means $m_{1,h} = -m_{1,l}$ assumed for symmetry and optimal detection. Each $s_1^n$ is real-valued and independently generated.

U$_2$ follows a variance modulation scheme, transmitting zero-mean Gaussian samples with two possible variances depending on the bit. That is, for bit-0 and bit-1, we have $s_2^n \sim \mathcal{N}(0, \sigma_{2,l}^2)$ and $s_2^n \sim \mathcal{N}(0, \sigma_{2,h}^2)$, with fixed mean 0 and variances satisfying $\sigma_{2,h}^2 = \alpha \sigma_{2,l}^2$, where $\alpha > 1$. Here, $\alpha$ represents the ratio between the high and low variance levels used by U$_2$. Unlike phase and amplitude modulations, variance modulation encodes information in the power level of noise rather than its shape or position.

U$_3$ modulates the correlation structure between pairs of noise samples to encode its bits. Each bit period consists of $N$ real samples grouped into $N/2$ correlated pairs. The samples are jointly Gaussian with a non-zero mean vector 
\(
\mathbf{m}_3 = [m_3,\, m_3]^\top
\) and a covariance matrix determined by the information bit:

\[
\mathbf{s}_{3}^n \sim \mathcal{N}(\mathbf{m}_3, \mathbf{\Sigma}_k), \quad
\mathbf{\Sigma}_k = \sigma_3^2
\begin{bmatrix}
1 & \rho_k \\
\rho_k & 1
\end{bmatrix}, \quad k \in \{l, h\}
\]

where \( \rho_l \) and \( \rho_h \) correspond to bit-0 and bit-1, respectively. Here, \( b_1 \), \( b_2 \), and \( b_3 \) represent the information bits of U$_1$, U$_2$, and U$_3$, respectively. Specifically, \( b_1 \) is conveyed by modulating the mean of the Gaussian noise samples, \( b_2 \) by modulating the variance, and \( b_3 \) by modulating the correlation coefficient between two jointly Gaussian noise samples.

We assume an average transmit power of $P$ for each user, which corresponds to the second moment of the transmitted samples. For U$_1$, this leads to $P = m_{1,l}^2 + \sigma_1^2$. A power allocation parameter $\beta \in (0,1)$ is defined such that $\sigma_1^2 = \beta P$ and $m_{1,l}^2 = (1 - \beta)P$. For U$_2$, assuming equiprobable bits, the average power is given by $P = (\sigma_{2,l}^2 + \sigma_{2,h}^2)/2$. For U$_3$, the power constraint is satisfied by keeping the marginal variances fixed at $\sigma_3^2$, such that each individual sample still satisfies $\mathbb{E}[(s_3^n)^2] = P$.

To facilitate the analysis, we define $\delta = \sigma_{2,l}^2 / \sigma_w^2$ as the normalized low-variance parameter of U$_2$, and $\alpha = \sigma_{2,h}^2 / \sigma_{2,l}^2$ as the high-to-low variance ratio. Additionally, the normalized variance of U$_1$ becomes $\eta = \sigma_1^2 / \sigma_w^2 = \frac{(1 + \alpha) \delta \beta}{2}$.

The $n$th received sample at the base station is expressed as
\begin{equation}
y^n = h_1 s_1^n + h_2 s_2^n + h_3 s_3^n + w^n, \quad n = 1, \ldots, N,
\label{eq:system_model_3user}
\end{equation}
where $h_1$, $h_2$, and $h_3$ denote the complex baseband fading coefficients between U$_1$, U$_2$, U$_3$ and the base station, respectively, and $w^n \sim \mathcal{CN}(0, \sigma_w^2)$ represents the additive white Gaussian noise (AWGN). We assume that our channel conditions are under the effects of Rician fading. Under Rician fading with factor \( K \), we have:
\begin{equation}
h_{i,R},\ h_{i,I} \sim \mathcal{N}\left( \sqrt{\frac{K}{2(1+K)}},\ \frac{1}{2(1+K)} \right), \quad i = 1, 2, 3.
\end{equation}
where \( R \) and \( I \) correspond to the real and imaginary components of the complex channel coefficient \( h_i \), respectively, and \( i \) denotes the user index in the system.

Conditioned on the transmitted bits and channel coefficients, the received signal has the following mean and variance:
\begin{equation}
\begin{aligned}
\mathrm{E}[y^n] &= h_1 m_{1,i} + h_3 m_3, \\
\mathrm{VAR}[y^n] &= |h_1|^2 \sigma_1^2 + |h_2|^2 \sigma_{2,k}^2 + |h_3|^2 \sigma_3^2 + \sigma_w^2,
\end{aligned}
\end{equation}
where $i,k \in \{l, h\}$ refer to the mean and variance bit indices of U$_1$ and U$_2$, respectively.

The base station (BS) jointly receives all users' signals over different channels and must extract each bit via statistical inference on the received samples. The next sections present the detection strategies and theoretical bit error probability (BEP) derivations for all three users.

\subsection{Uplink – User 1 Detection}

In the three-user ND-NOMA uplink system, the BS detects $\text{U}_1$'s bit by leveraging the fact that $\text{U}_1$ uses mean modulation. That is, each bit is transmitted over $N$ independent real Gaussian samples whose mean alternates depending on the bit value, while the variance remains fixed. To retrieve this embedded information, the BS computes the empirical mean of the received signal across the $N$ samples:
\begin{equation}
\bar{y} = \frac{1}{N} \sum_{n=1}^{N} y^n,
\end{equation}
where $y^n$ denotes the received signal at time index $n$. Due to the presence of multiple users and additive noise, $\bar{y}$ is a complex-valued random variable following a Gaussian distribution whose mean and variance are influenced by all three users’ transmission parameters and the instantaneous channel realizations.

Specifically, the conditional mean of $\bar{y}$, given the channel coefficients and transmitted bits, is expressed as $\mathbb{E}[\bar{y}] = h_1 m_{1,i} + h_3 m_3$, where $m_{1,i}$ represents the bit-dependent mean value transmitted by $\text{U}_1$ and $m_3$ is the constant mean of the samples transmitted by $\text{U}_3$. Here, $h_1$ and $h_3$ denote the complex baseband fading coefficients between the BS and U$_1$ and U$_3$, respectively. The term $h_1 m_{1,i}$ reflects the desired signal component, while $h_3 m_3$ acts as structured interference due to the non-zero mean of $\text{U}_3$’s signal.

The variance of $\bar{y}$, conditioned on the transmitted bits, is given by
\begin{equation}
\mathrm{VAR}[\bar{y}] = \frac{1}{N} \left( |h_1|^2 \sigma_1^2 + |h_2|^2 \sigma_{2,k}^2 + \eta P |h_3|^2 + \sigma_w^2 \right),
\end{equation}
where $\sigma_1^2$ is the fixed variance used by $\text{U}_1$, and $\sigma_{2,k}^2$ is the bit-dependent variance of the zero-mean samples sent by $\text{U}_2$, with $k \in \{l,h\}$ indicating low or high variance depending on the bit. The parameter $\eta P$ represents the average power allocated to $\text{U}_3$’s correlation-modulated signal, which contributes additional variability to the received signal. Finally, $\sigma_w^2$ is the variance of the AWGN at the BS.

Given this statistical structure, the BS applies a minimum-distance decision rule to determine $\text{U}_1$’s transmitted bit. The detector compares the received mean to two hypotheses, each corresponding to one of the possible transmitted means:
\begin{equation}
\hat{b}_1 = \begin{cases}
0, \text{if } |\bar{y} - h_1 m_{1,l} - h_3 m_3|^2 < |\bar{y} - h_1 m_{1,h} - h_3 m_3|^2, \\
1, \text{if } |\bar{y} - h_1 m_{1,h} - h_3 m_3|^2 < |\bar{y} - h_1 m_{1,l} - h_3 m_3|^2.
\end{cases}
\end{equation}

This rule corresponds to a minimum distance detector applied in the mean domain. The detector compares the sample mean of the received signal with two predefined reference means, each representing one of the two possible transmitted symbols. These reference means incorporate the average power contributions from the other two users, making the detector resilient to their interference without requiring explicit cancellation.

Under the assumption of symmetry between bit-0 and bit-1, and without loss of generality, we consider the probability of error when bit-0 is transmitted. Letting $m_{1,h} = -m_{1,l}$ for analytical simplicity, the conditional BEP becomes
\begin{equation}
\begin{aligned}
P_b = P\Big(&|\bar{y} - h_1 m_{1,h} - h_3 m_3|^2 \\
           &< |\bar{y} - h_1 m_{1,l} - h_3 m_3|^2 \mid b_1 = 0 \Big).
\end{aligned}
\end{equation}

This expression is further simplified by defining a decision variable $D = \mathrm{Re}\{ (\bar{y} - h_3 m_3) h_1^* m_{1,l} \}$, which is a real-valued Gaussian random variable with mean $
m_D=\mathbb{E}[D] = |h_1|^2 m_{1,l}^2$ and variance $\mathrm{VAR}[D] = \sigma_D^2$. The BEP is thus rewritten in $Q$-function form as
\begin{equation}
P_b = Q\left( \frac{m_D}{\sigma_D} \right)
\end{equation}
where the variance $\sigma_D^2$ captures the uncertainty in the detection variable and is given by
\begin{align}
\sigma_D^2 &= m_{1,l}^2 \left( h_{1,R}^2 \mathrm{VAR}[\bar{y}_R] + h_{1,I}^2 \mathrm{VAR}[\bar{y}_I] \right. \nonumber \\
& \quad \left. + 2 h_{1,R} h_{1,I} \mathrm{COV}(\bar{y}_R, \bar{y}_I) \right),
\end{align}
 $\bar{y}_R$, $\bar{y}_I$ being the corresponding components of $\bar{y}$. The individual variances and covariances in this expression are computed as follows:
 
\begin{equation}
\begin{aligned}
\mathrm{VAR}[\bar{y}_R] &= \frac{1}{N} \left( h_{1,R}^2 \sigma_1^2 + h_{2,R}^2 \sigma_{2,k}^2 + \eta P h_{3,R}^2 + \frac{\sigma_w^2}{2} \right), \\
\mathrm{VAR}[\bar{y}_I] &= \frac{1}{N} \left( h_{1,I}^2 \sigma_1^2 + h_{2,I}^2 \sigma_{2,k}^2 + \eta P h_{3,I}^2 + \frac{\sigma_w^2}{2} \right), \\
\mathrm{COV}(\bar{y}_R, \bar{y}_I) &= \frac{1}{N} \Big( h_{1,R} h_{1,I} \sigma_1^2 + h_{2,R} h_{2,I} \sigma_{2,k}^2 \\
&\quad + \eta P h_{3,R} h_{3,I} \Big)
\end{aligned}
\end{equation}

Since this BEP expression depends on the random fading coefficients $h_1$, $h_2$, $h_3$, and the bit-dependent variance $\sigma_{2,k}^2$, we average over all distributions to obtain the unconditional BEP. Respectively, we obtain
\begin{IEEEeqnarray}{rCl}
\bar{P}_b &=& \int \!\! \int \!\! \int \!\! \int \!\! \int \!\! \int
P_b \cdot f(h_{1,R}) f(h_{1,I}) f(h_{2,R}) f(h_{2,I}) \nonumber \\
&& \cdot f(h_{3,R}) f(h_{3,I}) \, dh_{1,R} \, dh_{1,I} \, dh_{2,R} \, dh_{2,I} \nonumber \\
&& \cdot \, dh_{3,R} \, dh_{3,I},
\label{eq:ubep}
\end{IEEEeqnarray}
where averaging is also performed over the two equiprobable values of $\sigma_{2,k}^2$. Since this integral is analytically intractable due to the complex dependencies within the $Q$ function, we employ Monte Carlo integration method, as detailed in Appendix~A, to numerically estimate the unconditional BEP.

\subsection{Uplink – User 2 Detection}

Unlike the  $\text{U}_1$’s mean detection scheme, this time BS involves a variance-based detection process in order to detect the bit of $\text{U}_2$. In this detection setting, the BS aims to detect whether $\text{U}_2$ has sent a Gaussian signal with low or high variance by observing the total energy of the received signal. While the decision process takes place in the real and positive domain, deriving an exact theoretical model is difficult as the simultaneous presence of three overlapping signals, each influenced by complex fading, noise, and especially the correlation-shaped samples originating from $\text{U}_3$.

To begin, the BS computes the sample variance of the received signal using
\begin{equation}
s_y^2 = \frac{1}{N-1} \sum_{n=1}^{N} \left| y^n - \bar{y} \right|^2,
\label{eq:sample_variance}
\end{equation}
where $y^n$ is the $n$th complex-valued received sample and $\bar{y}$ is the sample mean. 

The conditional variances of $y^n$, corresponding to bit-0 and bit-1 of $\text{U}_2$, are given by:
\begin{equation}
\begin{aligned}
s_0^2 &= \sigma_1^2 |h_1|^2 + \sigma_{2,l}^2 |h_2|^2 + \eta P |h_3|^2 + \sigma_w^2, \\
s_1^2 &= \sigma_1^2 |h_1|^2 + \sigma_{2,h}^2 |h_2|^2 + \eta P |h_3|^2 + \sigma_w^2
\end{aligned}
\end{equation}

The BS applies a decision rule that compares the observed sample variance $s_y^2$ with a reference threshold $\gamma$, which is chosen to balance the detection likelihoods under both bit hypotheses. This threshold can be computed optimally with the given conditional variances $s_0^2$ and $s_1^2$ as \cite{OPTIMUM_VAR_ML_THRESHOLD_2024}

\begin{equation}
    \gamma = \ln{\left( \frac{s_1^2}{s_0^2}\right)}\frac{s_1^2 \, s_0^2}{s_1^2 - s_0^2}.
\end{equation}

Then, the detection rule is:
\begin{equation}
\hat{b}_2 = \begin{cases}
0, & \text{if } s_y^2 < \gamma, \\
1, & \text{if } s_y^2 >\gamma.
\end{cases}
\end{equation}

The conditional bit error probability (BEP) of $\text{U}_2$ is expressed as:
\begin{equation}
P_b = \frac{1}{2} P(s_y^2 > \gamma \mid b_2 = 0) + \frac{1}{2} P(s_y^2 < \gamma \mid b_2 = 1).
\end{equation}

To analytically approximate this BEP for large $N$, we replace $\bar{y}$ in \eqref{eq:sample_variance} with its expected value $\mathbb{E}[\bar{y}] = h_1 m_{1,i} + h_3 m_3$, yielding the approximation:

\begin{align}
s_y^2 &\approx \frac{1}{N-1} \sum_{n=1}^{N} \left| y^n - h_1 m_{1,i} - h_3 m_3 \right|^2 \nonumber \\
&= \sum_{n=1}^{N} \Big[ 
    \left( y^n_R - h_{1,R} m_{1,i} - h_{3,R} m_3 \right)^2 \nonumber \\
&\quad + \left( y^n_I - h_{1,I} m_{1,i} - h_{3,I} m_3 \right)^2 
\Big].
\end{align}

By stacking the real and imaginary parts into a $2N \times 1$ real vector $\mathbf{y}$, this expression can be rewritten as a quadratic form:
\begin{equation}
s_y^2 = \mathbf{y}^T \Lambda \mathbf{y},
\end{equation}
where $\mathbf{y} \sim \mathcal{N}(0, \Sigma)$ and $\Lambda$ is a diagonal matrix. Here, 
$c$ denotes the covariance between the real and imaginary components of the received signal. The covariance matrix $\Sigma$ includes real-valued variances and off-diagonal correlation terms between the real and imaginary components due to channel fading:
\begin{equation}
\Sigma = \begin{bmatrix}
\sigma_R^2 & c & 0 & \cdots & 0 \\
c & \sigma_I^2 & 0 & \cdots & 0 \\
0 & 0 & \ddots & & \vdots \\
\vdots & \vdots & & \sigma_R^2 & c \\
0 & 0 & \cdots & c & \sigma_I^2
\end{bmatrix}_{2N \times 2N},
\end{equation}

Here, \( \epsilon \triangleq \eta \cdot \frac{P}{\sigma_w^2} \) denotes the normalized power contribution of User-3's correlation-modulated signal relative to the noise power.
\begin{equation}
\begin{aligned}
\sigma_R^2 &= h_{1,R}^2 \sigma_1^2 + h_{2,R}^2 \sigma_{2,k}^2 + \epsilon \sigma_w^2 h_{3,R}^2 + \frac{\sigma_w^2}{2}, \\
\sigma_I^2 &= h_{1,I}^2 \sigma_1^2 + h_{2,I}^2 \sigma_{2,k}^2 + \epsilon \sigma_w^2 h_{3,I}^2 + \frac{\sigma_w^2}{2}, \\
c &= h_{1,R} h_{1,I} \sigma_1^2 + h_{2,R} h_{2,I} \sigma_{2,k}^2 + \epsilon \sigma_w^2 h_{3,R} h_{3,I}
\end{aligned}
\end{equation}

Assuming $N$ is large enough, we invoke the central limit theorem and approximate $s_y^2 \sim \mathcal{N}(\mu_s, \sigma_s^2)$, where
\begin{equation}
\begin{aligned}
\mu_s &= \frac{N}{N-1} (\sigma_R^2 + \sigma_I^2), \\
\sigma_s^2 &= \frac{2N}{(N-1)^2} \left( \sigma_R^4 + \sigma_I^4 + 2c^2 \right)
\end{aligned}
\end{equation}

Substituting into the BEP formula yields:
\begin{equation}
P_b = \frac{1}{2} Q\left( \frac{\gamma - \mu_{s|b_2=0}}{\sigma_{s|b_2=0}} \right) + \frac{1}{2} Q\left( \frac{\mu_{s|b_2=1} - \gamma}{\sigma_{s|b_2=1}} \right).
\end{equation}
The threshold for equal error probabilities is
\begin{equation}
\gamma = \frac{ \sigma_{s|b_2=0} \, \mu_{s|b_2=1} \,+\, \sigma_{s|b_2=1} \, \mu_{s|b_2=0} }{ \sigma_{s|b_2=0} \,+\, \sigma_{s|b_2=1} },
\end{equation}
leading to a simplified BEP expression:
\begin{equation}
P_b = Q\left( \frac{\mu_{s|b_2=1} - \mu_{s|b_2=0}}{\sigma_{s|b_2=0} + \sigma_{s|b_2=1}} \right).
\end{equation}

As with $\text{U}_1$, the conditional BEP of $\text{U}_2$ decays with increasing $\sqrt{N}$ due to the averaging in the variance estimate. However, due to the presence of multiple interfering signals and nonlinear dependence on channel coefficients, a closed-form expression for the unconditional BEP is not tractable. Therefore, numerical integration over the distributions of $h_1$, $h_2$, and $h_3$ is used. Despite the mathematical complexity, the detection process remains computationally simple, requiring only variance computation and comparison with a fixed threshold, which is expectedly a simpler detection model than traditional SIC-based demodulation.

\subsection{Uplink – User 3 Detection}

In the proposed three-user ND-NOMA uplink system, $\text{U}_3$ encodes its information bits by modulating the correlation between consecutive Gaussian samples. This method introduces a new dimension of modulation, referred to as correlation modulation, and allows $\text{U}_3$ to communicate independently of the mean and variance dimensions used by $\text{U}_1$ and $\text{U}_2$.

For each bit duration, $\text{U}_3$ transmits $N$ real-valued Gaussian samples that are grouped into $N/2$ correlated pairs. These sample pairs are jointly drawn from a bivariate normal distribution with a specified correlation coefficient $\rho_k \in \{\rho_l, \rho_h\}$, corresponding to bit-0 and bit-1, respectively.

At the base station, the received signal is denoted as $\mathbf{y} = [y^1, y^2, \ldots, y^N]$, and is composed of contributions from all three users and noise. To recover $\text{U}_3$'s bit, the base station divides $\mathbf{y}$ into two equal-length segments:
\begin{equation}
\mathbf{y}_1 = [y^1, \ldots, y^{N/2}], \quad \mathbf{y}_2 = [y^{N/2+1}, \ldots, y^N],
\end{equation}
and calculates the empirical covariance between them:
\begin{equation}
\hat{\rho}_y = \frac{2}{N} \sum_{n=1}^{N/2} \mathrm{Re} \left\{ (y^n - \bar{y}_1)(y^{n + N/2} - \bar{y}_2)^* \right\},
\end{equation}
where $\bar{y}_1$ and $\bar{y}_2$ are the respective sample means of $\mathbf{y}_1$ and $\mathbf{y}_2$. This estimator targets the underlying second-order correlation imposed by $\text{U}_3$'s transmitted noise structure.

The base station then compares $\hat{\rho}_y$ to the expected correlation values under each hypothesis:
\begin{equation}
\begin{aligned}
\hat{\rho}_l &= |h_3|^2 \rho_l \eta P, \\
\hat{\rho}_h &= |h_3|^2 \rho_h \eta P,
\end{aligned}
\end{equation}
where $\eta P$ represents the power allocated by $\text{U}_3$ and $\rho_l$, $\rho_h$ are the correlation values used to modulate bit-0 and bit-1. The detection rule is a minimum distance test:

\begin{equation}
\hat{b}_3 = \begin{cases}
0, & \text{if } |\hat{\rho}_y - \hat{\rho}_l|^2 < |\hat{\rho}_y - \hat{\rho}_h|^2, \\
1, & \text{if } |\hat{\rho}_y - \hat{\rho}_l|^2 > |\hat{\rho}_y - \hat{\rho}_h|^2.
\end{cases}
\end{equation}

To characterize the BEP, we define a Gaussian random variable $D = \hat{\rho}_y - (\hat{\rho}_l + \hat{\rho}_h)/2$, and under the assumption of large $N$, $D$ is approximately Gaussian due to averaging. Thus, the conditional BEP is written as:
\begin{equation}
P_b = Q\left( \frac{|\hat{\rho}_h - \hat{\rho}_l|}{2 \sigma_C} \right),
\end{equation}
where $\sigma_C^2$ is the variance of $\hat{\rho}_y$ and captures the uncertainty due to both interference and noise. Specifically, it includes contributions from all three users. Here, \( \sigma_{1,C}^2 \), \( \sigma_{2,C}^2 \), and \( \sigma_{w,C}^2 \) denote the contributions of U\(_1\)’s mean-modulated symbols, U\(_2\)’s variance modulation, and the additive channel noise, respectively, to the variance of the correlation estimate \( \hat{\rho}_y \).: 
\begin{equation}
\sigma_C^2 \approx \frac{1}{N/2} \left( \sigma_{1,C}^2 + \sigma_{2,C}^2 + \sigma_{w,C}^2 \right).
\end{equation}
These terms represent the impact of $\text{U}_1$’s mean-modulated samples, $\text{U}_2$’s variance modulation, and the additive channel noise. In practice, these variances are computed numerically in simulation using the channel realizations and transmitted bits.

Due to the complexity of closed-form expressions, the integral of the unconditional BEP is once again evaluated via Monte Carlo integration, as outlined in Appendix~A. As observed in $\text{U}_1$ and $\text{U}_2$ detection, the conditional BEP of $\text{U}_3$ also decays with increasing $N$ due to the law of large numbers, improving the accuracy of the empirical covariance estimation.

\begin{figure*}[h]
    \centering
    \includegraphics[width=1\linewidth]{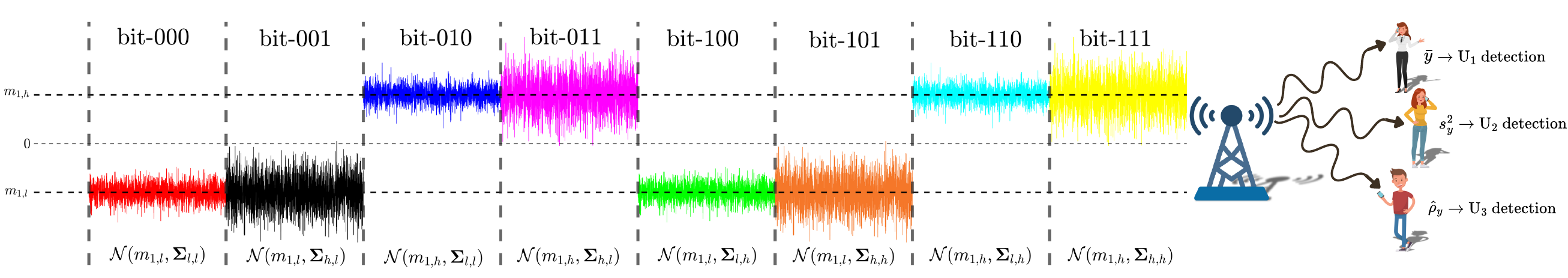}
    \caption{Downlink Three-User ND-NOMA scheme using real Gaussian signals.}
    \label{fig:downlink}
\end{figure*}

\section{Downlink Three-User ND-NOMA: System Model and
Performance Analysis}
In this section, we present the downlink architecture of three-user ND-NOMA, where user data is simultaneously embedded into the mean, variance, and correlation of Gaussian noise samples.

\subsection{System Model}

As shown in Fig.~\ref{fig:downlink}, for the three-user ND-NOMA downlink scheme, the BS simultaneously serves three users by embedding their bits into different statistical domains of a common Gaussian noise signal. Specifically, $\text{U}_1$'s bit is mapped to the mean of the noise samples, $\text{U}_2$'s bit is encoded in the variance, and $\text{U}_3$'s bit is represented through the correlation between consecutive sample pairs. Binary modulation is considered among each of the users. The complete statistical characterization of the transmitted and received signals for each possible bit combination in the downlink three-user ND-NOMA system is summarized in Table~\ref{tab:3user_downlink_stats}.

\begin{table}[t]
\centering
\caption{Statistics of the transmitted and received signals for downlink three-user ND-NOMA}
\begin{tabular}{|c|c|c|}
\hline
U$_1$/U$_2$/U$_3$ bits & {$s^n_{BS} \sim$} & $y_i^n \sim$ for $i = 1,2,3$ \\
\hline
000 & $\mathcal{N}(m_l, \mathbf{\Sigma}_l,_l)$ & $\mathcal{CN}(h_i m_l, |h_i|^2 \sigma^2_{2,l} + \sigma_w^2)$ \\ \hline
001 & $\mathcal{N}(m_l, \mathbf{\Sigma}_h,_l)$ & $\mathcal{CN}(h_i m_l, |h_i|^2 \sigma^2_{2,h} + \sigma_w^2)$ \\ \hline
010 & $\mathcal{N}(m_l,  \mathbf{\Sigma}_l,_l)$ & $\mathcal{CN}(h_i m_l, |h_i|^2 \sigma^2_{2,l} + \sigma_w^2)$ \\ \hline
011 & $\mathcal{N}(m_l,  \mathbf{\Sigma}_h,_l)$ & $\mathcal{CN}(h_i m_l, |h_i|^2 \sigma^2_{2,h} + \sigma_w^2)$ \\ \hline
100 & $\mathcal{N}(m_h, \mathbf{\Sigma}_l,_h)$ & $\mathcal{CN}(h_i m_h, |h_i|^2 \sigma^2_{2,l} + \sigma_w^2)$ \\ \hline
101 & $\mathcal{N}(m_h,  \mathbf{\Sigma}_h,_h)$ & $\mathcal{CN}(h_i m_h, |h_i|^2 \sigma^2_{2,h} + \sigma_w^2)$ \\ \hline
110 & $\mathcal{N}(m_h,  \mathbf{\Sigma}_l,_h)$ & $\mathcal{CN}(h_i m_h, |h_i|^2 \sigma^2_{2,l} + \sigma_w^2)$ \\ \hline
111 & $\mathcal{N}(m_h,  \mathbf{\Sigma}_h,_h)$ & $\mathcal{CN}(h_i m_h, |h_i|^2 \sigma^2_{2,h} + \sigma_w^2)$ \\ \hline
\end{tabular}
\label{tab:3user_downlink_stats}
\end{table}

The BS transmits a sequence of $N$ real-valued Gaussian noise samples per bit duration, with distribution defined as:
\begin{equation}
s_{BS}^n \sim \mathcal{N}(m_{1,i}, \boldsymbol{\mathbf{\Sigma}}_{k,j}), \quad i,j,k \in \{l,h\},
\end{equation}
where $m_{1,i}$ is the bit-dependent mean associated with U$_1$ (bit $\mathrm{b}_{1,i}$),  
and $\sigma_{k}^2$ is the variance corresponding to U$_2$’s bit $\mathrm{b}_{2,k}$.  
To embed U$_3$’s information, we group the $N$ real-valued samples into $N/2$ bivariate pairs.  
Each pair $(s_{\mathrm{BS}}^{2n-1},\,s_{\mathrm{BS}}^{2n})$ uses a bit-dependent correlation coefficient  
$\rho_j\in\{\rho_{l},\rho_{h}\}$. To jointly embed the bits of U$_2$ and U$_3$ in the noise domain, we define the combined covariance matrix as:
\begin{equation}
\boldsymbol{\Sigma}_{k,j} = \sigma_{2,k}^2 
\begin{bmatrix}
1 & \rho_j \\
\rho_j & 1
\end{bmatrix}, \quad k, j \in \{l, h\}
\end{equation}
This formulation captures the bit-dependent power allocation of U$_2$ through the variance term \(\sigma_{2,k}^2\), and the bit-dependent correlation imposed by U$_3$ via \(\rho_j\), resulting in a noise structure that encodes both users' information simultaneously.

The resulting noise sequence, jointly modulated in mean, variance, and correlation, is transmitted through fading channels. The received signals at the three users are:
\begin{subequations}
\begin{align}
y_1^n &= h_1 s_{BS}^n + w_1^n, \tag{\theequation.a} \\
y_2^n &= h_2 s_{BS}^n + w_2^n, \tag{\theequation.b} \\
y_3^n &= h_3 s_{BS}^n + w_3^n, \tag{\theequation.c} \label{eq:downlink_c}
\end{align}
\end{subequations}
for $n = 1, \ldots, N$. Here, $h_1$, $h_2$, and $h_3$ denote the complex downlink channel fading coefficients between the BS and each user, and $w_i^n \sim \mathcal{CN}(0, \sigma_w^2)$ is the AWGN at $\text{U}_i$ for $i = 1,2,3$.

We assume the BS transmits with a total power budget of $P$, such that:
\begin{equation}
\mathbb{E}[(s_{BS}^n)^2] = P.
\end{equation}
To divide this power among the three users, we define a power partitioning parameter $\psi$ for $\text{U}_1$ and $(1-\psi)$ for the remaining variance and correlation modulation. More specifically, the mean of the signal is allocated a portion $\psi$ of the total power (DC component), and the variance, which supports both $\text{U}_2$ and $\text{U}_3$, receives $(1-\psi)P$ (AC component). The bit-dependent means satisfy $m_{1,h} = -m_{1,l}$ and $m_{1,l}^2 = \psi P$, while the variances are chosen such that:
\begin{equation}
\frac{\sigma_{2,l}^2 + \sigma_{2,h}^2}{2} = (1 - \psi)P, \quad \sigma_{2,h}^2 = \alpha \sigma_{2,l}^2.
\end{equation}

To facilitate correlation modulation, each pair of noise samples has a correlation coefficient $\rho_j$ determined by $\text{U}_3$’s bit. These correlation values are selected such that the second moment of each sample remains within the overall power constraint, preserving the total energy allocated to the ac component. 

Unlike in uplink ND-NOMA, the downlink signals do not overlap in terms of physical transmission — a single superimposed signal is broadcast by the BS. However, since each user extracts their bit from a different statistical property of the signal, there is no need for SIC, and detection can be performed independently at each receiver using straightforward statistical operations. As a result, the downlink ND-NOMA system benefits from reduced complexity and avoids typical error propagation problems encountered in PD-NOMA systems.

In the subsequent sections, we detail the detection mechanisms used by each user to extract their respective information from the received signal and provide analytical expressions for their corresponding BEP.

\subsection{Downlink – User 1 Detection}

In the downlink transmission of the proposed scheme, $\text{U}_1$ recovers its bit by exploiting the mean modulation embedded by the BS in the transmitted noise samples. This detection mechanism mirrors that of the uplink, but the composite interference now also includes contributions from both $\text{U}_2$'s variance modulation and $\text{U}_3$'s correlation-based structure.

$\bar{y}_R$ and $\bar{y}_I$ denote the real and imaginary components of $\bar{y}$, respectively.  Their individual variances and covariance are

\begin{equation}
\begin{aligned}
\mathrm{VAR}[\bar{y}_R] &= \frac{1}{N}\!\left( h_{1,R}^{2}\,\sigma_{1}^{2}
                     + h_{2,R}^{2}\,\sigma_{2,k}^{2}
                     + \frac{\beta P^{2}}{\sigma_{w}^{2}}\,h_{3,R}^{2}
                     + \frac{\sigma_{w}^{2}}{2} \right),\\[4pt]
\mathrm{VAR}[\bar{y}_I] &= \frac{1}{N}\!\left( h_{1,I}^{2}\,\sigma_{1}^{2}
                     + h_{2,I}^{2}\,\sigma_{2,k}^{2}
                     + \frac{\beta P^{2}}{\sigma_{w}^{2}}\,h_{3,I}^{2}
                     + \frac{\sigma_{w}^{2}}{2} \right),\\[4pt]
\mathrm{COV}[\bar{y}_R,\bar{y}_I] 
&= \frac{1}{N}\!\left( h_{1,R}h_{1,I}\,\sigma_{1}^{2}
                     + h_{2,R}h_{2,I}\,\sigma_{2,k}^{2} \right.\\
&\quad \left. + \frac{\beta P^{2}}{\sigma_{w}^{2}}\,h_{3,R}h_{3,I} \right),
\end{aligned}
\end{equation}

where $m_{1,i} \in \{m_{1,l}, m_{1,h}\}$ is the bit-dependent mean used by the BS to represent $\text{U}_1$'s bit, and $\sigma_{2,k}^2$ is the variance selected based on $\text{U}_2$'s bit. The term $\frac{\beta P^2}{\sigma_w^2}$ accounts for the power allocated to $\text{U}_3$'s correlation-modulated signal, which appears as structured interference at all users, scaled by the downlink channel coefficient $h_1$.

Given this distribution, the optimal minimum-distance detector for $\text{U}_1$ is expressed as:
\begin{equation}
\hat{b}_1 = \begin{cases}
0, & \text{if } |\bar{y}_1 - h_1 m_{1,l}|^2 < |\bar{y}_1 - h_1 m_{1,h}|^2, \\
1, & \text{if } |\bar{y}_1 - h_1 m_{1,l}|^2 > |\bar{y}_1 - h_1 m_{1,h}|^2.
\end{cases}
\end{equation}

This rule leads to the following expression for the conditional BEP:
\begin{align}
P_b &= P\left( |\bar{y}_1 - h_1 m_{1,h}|^2 < |\bar{y}_1 - h_1 m_{1,l}|^2 \mid b_1 = 0 \right) \nonumber \\
&= P\left( \mathrm{Re} \left\{ \bar{y}_1 h_1^* m_{1,l} \right\} < 0 \mid b_1 = 0 \right) \nonumber \\
&= Q\left( \frac{m_D}{\sigma_D} \right),
\end{align}
where $D = \mathrm{Re} \left\{ \bar{y}_1 h_1^* m_{1,l} \right\}$ is a real Gaussian variable with
\begin{equation}
\begin{aligned}
m_D &= |h_1|^2 m_{1,l}^2, \\
\sigma_D^2 &= m_{1,l}^2 \left( h_{1,R}^2 \mathrm{VAR}[\bar{y}_{1,R}] + h_{1,I}^2 \mathrm{VAR}[\bar{y}_{1,I}] \right. \\
&\quad \left. + 2 h_{1,R} h_{1,I} \mathrm{COV}(\bar{y}_{1,R}, \bar{y}_{1,I}) \right)
\end{aligned}
\end{equation}

The variances and covariance of the real and imaginary parts of $\bar{y}_1$ are given by:
\begin{equation}
\begin{aligned}
\mathrm{VAR}[\bar{y}_{1,R}] &= \frac{1}{N} \left( h_{1,R}^2 \sigma_{2,k}^2 + \frac{\beta P^2}{\sigma_w^2} h_{1,R}^2 + \frac{\sigma_w^2}{2} \right), \\
\mathrm{VAR}[\bar{y}_{1,I}] &= \frac{1}{N} \left( h_{1,I}^2 \sigma_{2,k}^2 + \frac{\beta P^2}{\sigma_w^2} h_{1,I}^2 + \frac{\sigma_w^2}{2} \right), \\
\mathrm{COV}(\bar{y}_{1,R}, \bar{y}_{1,I}) &= \frac{1}{N} \left( h_{1,R} h_{1,I} \sigma_{2,k}^2 + \frac{\beta P^2}{\sigma_w^2} h_{1,R} h_{1,I} \right)
\end{aligned}
\end{equation}

Substituting these into the expression for $\sigma_D^2$ yields the full conditional BEP for $\text{U}_1$.

Finally, as in the uplink case, the unconditional BEP is obtained via numerical averaging over the fading distribution of $h_1$, as well as over the equiprobable values of $\sigma_{2,k}^2$. The resulting performance illustrates that, despite the added interference from $\text{U}_3$, $\text{U}_1$’s detection remains efficient due to its orthogonal mapping into the mean of the signal.

\subsection{Downlink – User 2 Detection}

In the downlink phase of the three-user ND-NOMA system, $\text{U}_2$ extracts its bit from the variance of the received signal, in a fashion similar to the uplink detection scheme. However, the downlink environment introduces a different interference pattern since the transmitted signal now embeds the bits of all users in a single waveform. While $\text{U}_2$ still detects variance changes, its observation is affected by both $\text{U}_1$'s mean modulation and $\text{U}_3$'s correlation modulation.

To begin, the receiver at $\text{U}_2$ computes the sample variance of the received signal over $N$ samples:
\begin{equation}
s_{y_2}^2 = \frac{1}{N-1} \sum_{n=1}^{N} \left| y_2^n - \bar{y}_2 \right|^2,
\end{equation}
where $\bar{y}_2 = \frac{1}{N} \sum_{n=1}^{N} y_2^n$ is the sample mean. Based on the signal structure, the received signal is impacted by the mean and variance chosen by the BS according to $\text{U}_1$ and $\text{U}_2$'s bits, and further distorted by $\text{U}_3$'s correlation-induced sample structure. This results in conditional variance expressions:
\begin{equation}
\begin{aligned}
s_0^2 &= |h_2|^2 \sigma_{2,l}^2 + |h_2|^2 (1 - \psi) P + \sigma_w^2, \\
s_1^2 &= |h_2|^2 \sigma_{2,h}^2 + |h_2|^2 (1 - \psi) P + \sigma_w^2,
\end{aligned}
\end{equation}
where $\sigma_{2,l}^2$ and $\sigma_{2,h}^2$ represent the bit-dependent variances for $\text{U}_2$, and $(1 - \psi) P$ denotes the average power contribution of $\text{U}_3$'s signal, which adds structured variance interference that scales with $|h_2|^2$.

To decide between the two variance hypotheses, $\text{U}_2$ applies a threshold test:
\begin{equation}
\hat{b}_2 = \begin{cases}
0, & \text{if } s_{y_2}^2 < \gamma, \\
1, & \text{if } s_{y_2}^2 > \gamma,
\end{cases}
\end{equation}
with threshold $\gamma$, which can be calculated optimally as \cite{OPTIMUM_VAR_ML_THRESHOLD_2024}
\begin{equation}
    \gamma = \ln{\left( \frac{s_1^2}{s_0^2}\right)}\frac{s_1^2 \, s_0^2}{s_1^2 - s_0^2}.
\end{equation}

The conditional BEP becomes:
\begin{equation}
P_b = \frac{1}{2} P(s_{y_2}^2 > \gamma \mid b_2 = 0) + \frac{1}{2} P(s_{y_2}^2 < \gamma \mid b_2 = 1).
\end{equation}

To analyze this statistically, we approximate $s_{y_2}^2 \approx \frac{1}{N-1} \sum_{n=1}^{N} \left| y_2^n - h_2 m_{1,i} \right|^2$, replacing $\bar{y}_2$ by its conditional expectation $\mathbb{E}[\bar{y}_2] = h_2 m_{1,i}$ as done in the uplink model. This lets us express $s_{y_2}^2$ as a quadratic form of a $2N \times 1$ real Gaussian vector:
\begin{equation}
s_{y_2}^2 = \mathbf{y}^T \Lambda \mathbf{y},
\end{equation}
where $\mathbf{y} \sim \mathcal{N}(0, \Sigma)$, and $\Lambda$ is a diagonal matrix. The covariance matrix $\Sigma$ remains banded and is updated for the current scenario:
\begin{equation}
\Sigma = \begin{bmatrix}
\sigma_R^2 & c & 0 & \cdots & 0 \\
c & \sigma_I^2 & 0 & \cdots & 0 \\
0 & 0 & \ddots & & \vdots \\
\vdots & \vdots & & \sigma_R^2 & c \\
0 & 0 & \cdots & c & \sigma_I^2
\end{bmatrix},
\end{equation}
with:
\begin{equation}
\begin{aligned}
\sigma_R^2 &= h_{2,R}^2 \sigma_{2,k}^2 \;+\; (1-\psi)P\, h_{2,R}^2 \;+\; \frac{\sigma_w^2}{2}, \\
\sigma_I^2 &= h_{2,I}^2 \sigma_{2,k}^2 \;+\; (1-\psi)P\, h_{2,I}^2 \;+\; \frac{\sigma_w^2}{2}, \\
c &= h_{2,R} h_{2,I} \sigma_{2,k}^2 \;+\; (1-\psi)P\, h_{2,R} h_{2,I}.
\end{aligned}
\end{equation}

for $k \in \{l,h\}$, depending on the bit transmitted.

Assuming sufficiently large $N$, we invoke the central limit theorem and approximate $s_{y_2}^2 \sim \mathcal{N}(\mu_s, \sigma_s^2)$ with:
\begin{equation}
\begin{aligned}
\mu_s &= \frac{N}{N-1} (\sigma_R^2 + \sigma_I^2), \\
\sigma_s^2 &= \frac{2N}{(N-1)^2} (\sigma_R^4 + \sigma_I^4 + 2c^2).
\end{aligned}
\end{equation}

Substituting into the BEP equation, we obtain:
\begin{equation}
P_b = \frac{1}{2} Q\left( \frac{\gamma - \mu_{s|b_2=0}}{\sigma_{s|b_2=0}} \right) + \frac{1}{2} Q\left( \frac{\mu_{s|b_2=1} - \gamma}{\sigma_{s|b_2=1}} \right).
\end{equation}

The optimal threshold $\gamma$ for equal error probabilities is:
\begin{equation}
\gamma = \frac{\sigma_{s|b_2=0} \mu_{s|b_2=1} + \sigma_{s|b_2=1} \mu_{s|b_2=0}}{\sigma_{s|b_2=0} + \sigma_{s|b_2=1}},
\end{equation}
and the simplified expression for BEP becomes:
\begin{equation}
P_b = Q\left( \frac{\mu_{s|b_2=1} - \mu_{s|b_2=0}}{\sigma_{s|b_2=0} + \sigma_{s|b_2=1}} \right).
\end{equation}

As with the uplink case, the conditional BEP decays with increasing $N$, and the unconditional BEP is computed via numerical integration over the fading coefficients $h_2$ and $\text{U}_3$’s interference profile. Despite the added complexity from $\text{U}_3$, the detection remains computationally light and does not require interference cancellation.

\subsection{Downlink – User 3 Detection}

In the three-user ND-NOMA downlink system, $\text{U}_3$ decodes its bit based on correlation modulation—mirroring the structure used in the uplink case. Specifically, the BS encodes $\text{U}_3$’s information into the correlation between consecutive noise samples. Each bit of $\text{U}_3$ is transmitted by shaping the joint distribution of $N$ real-valued noise samples into $N/2$ bivariate Gaussian pairs with low or high correlation, corresponding to bit-0 and bit-1, respectively.

The transmitted signal $s_{BS}^n \sim \mathcal{N}(m_{1,i}, \mathbf{\Sigma}_{k,j})$ is further structured such that each sample pair $(s_{BS}^{2n-1}, s_{BS}^{2n})$ follows a bivariate normal distribution with correlation coefficient $\rho_j \in \{\rho_l, \rho_h\}$ depending on the bit of $\text{U}_3$. The received signal at $\text{U}_3$ is modeled as it was in (\ref{eq:downlink_c}).

To extract the embedded bit through correlation, $\text{U}_3$ splits its received vector into two halves:
\[
\mathbf{y}_1 = [y_3^1, \ldots, y_3^{N/2}], \quad \mathbf{y}_2 = [y_3^{N/2+1}, \ldots, y_3^N],
\]
and computes the real part of the sample correlation estimate:
\begin{equation}
\hat{\rho}_y = \frac{2}{N} \sum_{n=1}^{N/2} \mathrm{Re} \left\{ (y_3^n - \bar{y}_1)(y_3^{n+N/2} - \bar{y}_2)^* \right\},
\end{equation}
where $\bar{y}_1$ and $\bar{y}_2$ denote the sample means of the respective halves. This statistic estimates the correlation introduced at the transmitter and distorted by fading and additive noise.

The BS's correlation structure results in two reference correlation levels at the receiver:
\begin{equation}
\begin{aligned}
\hat{\rho}_l &= |h_3|^2 \rho_l \,(1-\psi) P, \\
\hat{\rho}_h &= |h_3|^2 \rho_h \,(1-\psi) P.
\end{aligned}
\end{equation}

Based on these references, the detection rule is given by
\begin{equation}
\hat{b}_3 = \begin{cases}
0, & \text{if } |\hat{\rho}_y - \hat{\rho}_l|^2 < |\hat{\rho}_y - \hat{\rho}_h|^2, \\
1, & \text{if } |\hat{\rho}_y - \hat{\rho}_l|^2 > |\hat{\rho}_y - \hat{\rho}_h|^2.
\end{cases}
\end{equation}

Assuming $\hat{\rho}_y$ is approximately Gaussian due to averaging over a large number of samples, the conditional bit error probability becomes:
\begin{equation}
P_b = Q\left( \frac{|\hat{\rho}_h - \hat{\rho}_l|}{2 \sigma_C} \right),
\end{equation}
where $\sigma_C^2$ denotes the variance of $\hat{\rho}_y$. This variance captures the impact of $\text{U}_1$’s mean, $\text{U}_2$’s variance modulation, and noise, all passed through the same fading channel $h_3$. An analytical approximation for the covariance variance is given by:
\begin{equation}
\sigma_C^2 \approx \frac{1}{N/2} \left( \sigma_{1,C}^2 + \sigma_{2,C}^2 + \sigma_{w,C}^2 \right),
\end{equation}
where each term quantifies the covariance uncertainty due to mean interference (from $\text{U}_1$), variance interference (from $\text{U}_2$), and thermal noise.

Since a closed-form solution for the unconditional BEP is analytically intractable, as detailed in Appendix~A, integration over the Rician fading coefficients to estimate the unconditional BEP.

\section{Numerical Results}

\begin{table}[t!]
\centering
\caption{Simulation Parameters for 3-User ND-NOMA System}
\renewcommand{\arraystretch}{1.2}
\begin{tabular}{|p{3.5cm}|p{3.5cm}|}
\hline
\textbf{Parameter} & \textbf{Value} \\ \hline \hline
$\delta_\text{dB}$ &  [-30, 10] , [-40, 10] \\ \hline
$\alpha$       & 10 \\ \hline
$P_\text{{dBm}}$        & 40 \\ \hline
$\beta$        & 1/100 (UL), 1/1024 (DL) \\ \hline
$\rho$         & -1, +1 \\ \hline
$\psi$         & 0.5 \\ \hline
\end{tabular}
\label{tab:params_3user}
\end{table}

In this section, we examine the BER performance of the proposed three-user ND-NOMA system for both uplink and downlink scenarios. We assume that each user's wireless channel is subject to Rician fading. To obtain average BEP values across fading conditions, we adopt the Monte Carlo integration technique whose details are given in Appendix A. This approach enables accurate evaluation of the unconditional BEP for all users under realistic channel conditions. The simulation outcomes are compared with theoretical predictions to validate the effectiveness of the system. Simulation parameters used throughout the analysis are presented in Table~\ref{tab:params_3user}.

\subsection{Uplink Scenario}

\begin{figure}[t]
    \centering
    \includegraphics[width=1\linewidth]{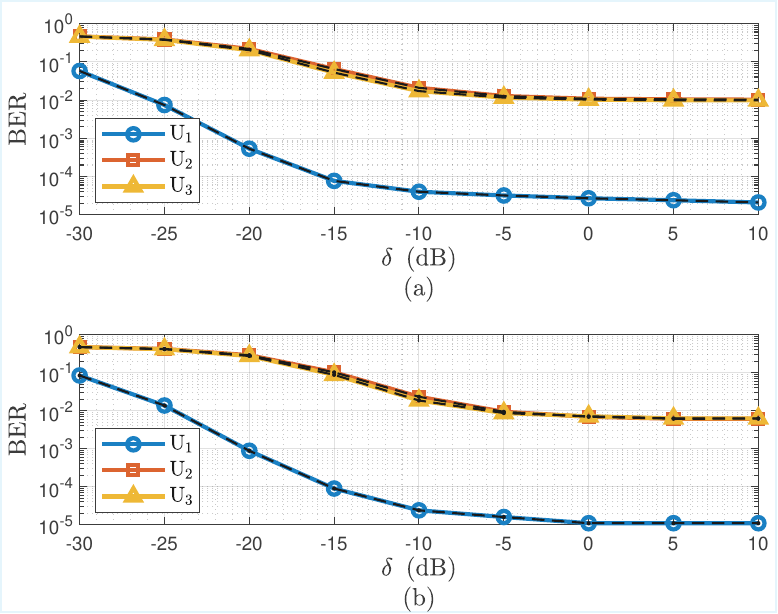}
    \caption{Simulated BER and theoretical BEP (shown as dashed black lines) for the uplink three-user ND-NOMA system versus $\delta$, under Rician fading with $K = 10$ dB. Results are shown for (a) $N = 150$ and (b) $N = 200$ samples.}
    
    \label{fig:fig3}
\end{figure}

\begin{figure}[t]
    \centering
    \includegraphics[width=1\linewidth]{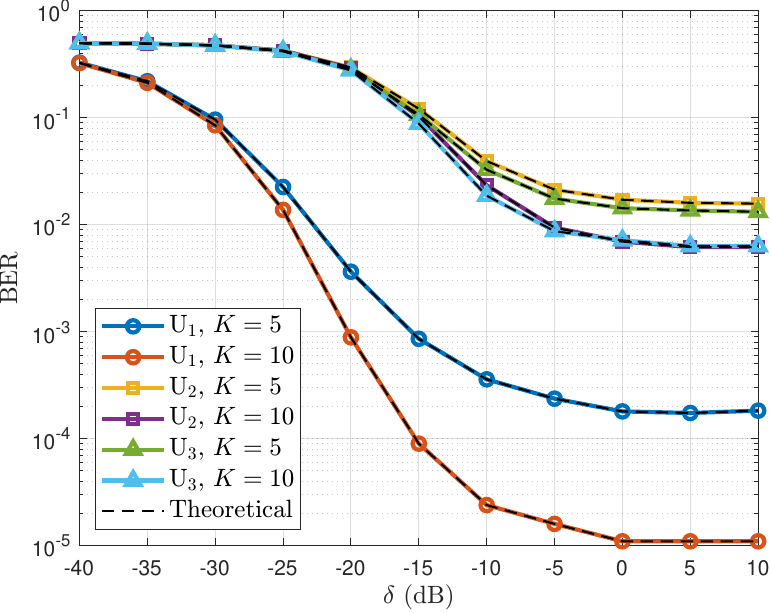}
    \caption{Theoretical BEP and simulated BER for uplink three-user ND-NOMA versus $\delta$ with Rician $K$-factors of 5 and 10.}
    \label{fig:fig4}
\end{figure}

Figs. \ref{fig:fig3} and \ref{fig:fig4} illustrate the uplink BER performance of U$_1$, U$_2$ and U$_3$. The BS jointly receives the superimposed signals from all users under a Rician fading channel, with varying noise configurations. Figs. \ref{fig:fig3}(a) and \ref{fig:fig3}(b) shows the BER performance of the three users for $N = 150$ and $N = 200$ samples, respectively, under a fixed Rician K-factor of $K = 10$ dB.

 It is observed that U$_1$ achieves the lowest BER among all users due to its use of mean modulation, which is more robust to noise and channel variations. On the other hand, U$_2$ and U$_3$, employing variance- and correlation-based modulation respectively, exhibit relatively higher BERs. This is mainly because their detection relies on more sensitive statistical features, which are more easily affected by interference and noise. Ultimately it must be stated that expectedly, increasing the number of noise samples from $N = 150$ to $N = 200$ improves BER performance for all users, as larger $N$ provides more reliable statistical estimates at the receiver.

As expected, increasing $\delta$ leads to improved BER performance for all users. A higher $\delta$ implies stronger signal components relative to noise, making it easier for the receiver to distinguish.

Fig.~\ref{fig:fig4} extends the analysis by comparing uplink BER performance for two different Rician $K$-factors, namely $K = 5$ dB and $K = 10$ dB where $N$ is fixed to 200. As expected, stronger line-of-sight conditions (higher $K$) lead to improved performance for all users, validating the analytical model's accuracy across channel conditions. The theoretical BEP expressions align well with simulation results, confirming the validity of our approximations.

\subsection{Downlink Scenario}

\begin{figure}[t]

    \centering
    \includegraphics[width=1\linewidth]{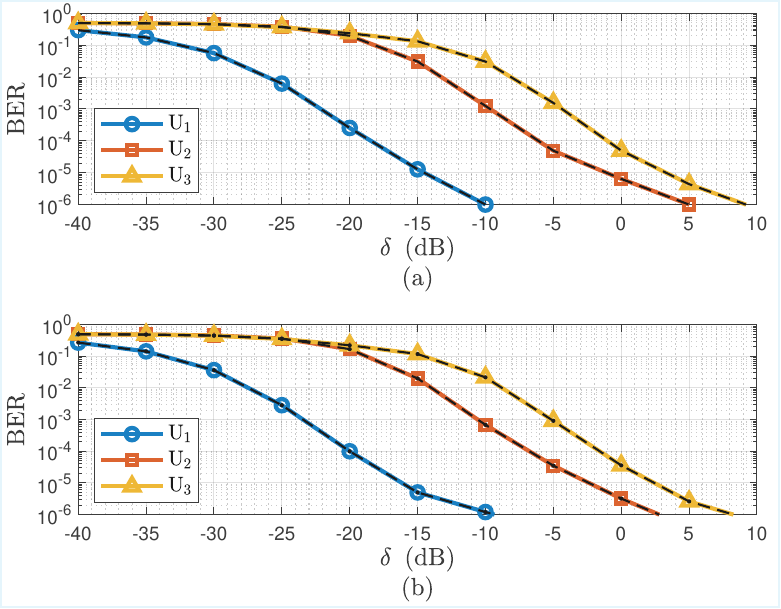}
    \caption{Simulated BER and theoretical BEP (shown as dashed black lines) for the downlink three-user ND-NOMA system versus $\delta$, under Rician fading with $K = 10$ dB. Results are shown for (a) $N = 150$ and (b) $N = 200$ samples.}
    \label{fig:fig5}
\end{figure}

\begin{figure}[t]
    \centering
    \includegraphics[width=0.49\textwidth]{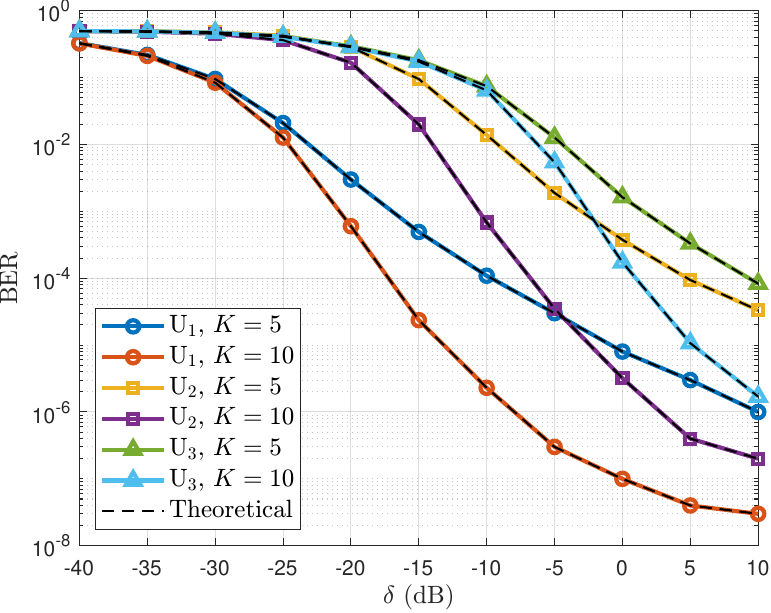}
    \caption{Theoretical BEP and simulated BER for downlink three-user ND-NOMA versus $\delta$ with Rician $K$-factors of 5 and 10.}
    \label{fig:fig6}
\end{figure}

Figs.~\ref{fig:fig5} and \ref{fig:fig6} illustrate the BER performance of U$_1$, U$_2$, and U$_3$ in the downlink three-user ND-NOMA system, where each user recovers their bit independently using simple detection rules. Fig.~\ref{fig:fig5}(a) and Fig. \ref{fig:fig5}(b) display BER results for $N=150$ and $N=200$ samples, respectively, while Fig.~\ref{fig:fig6} compares performance across different Rician $K$-factors, with $K \in \{5, 10\}$ dB at a fixed $N=200$.

As shown in Figs.~\ref{fig:fig5}(a) and \ref{fig:fig5}(b), U$_1$ achieves the best performance among all users just like the uplink scenario, benefiting from the robustness of mean-based detection under noise and fading. In contrast, U$_2$ and U$_3$ show relatively higher BERs due to their reliance on variance and correlation features, which are more susceptible to interference and statistical estimation error. Notably, U$_3$ slightly outperforms U$_2$ in the downlink case. This is attributed to the fact that correlation-based detection, while sensitive, benefits from the symmetrical structure of the broadcast signal, and offers simpler bit-level decisions based on a single statistic. Increasing the number of noise samples and $\delta$ improves BER performance for all users once again, as larger sample sizes yield more accurate statistical estimates.

Fig.~\ref{fig:fig6} further supports this by demonstrating that a higher Rician $K$-factor improves performance for all users. With $K = 10$ dB providing stronger line-of-sight conditions than $K = 5$ dB, resulting in lower BER. The exact match between theoretical and simulated curves confirms the accuracy of the analytical model in characterizing the downlink system. Furthermore, the satisfactory BER performance observed across users in the downlink scenario stems from the absence of mutual interference, as each user's information is embedded in orthogonal statistical domains, enabling reliable detection without the need of complex receiver processing.

\subsection{Comparison with PD-NOMA}

\begin{figure}[t]
    \centering
    \includegraphics[width=0.49\textwidth]{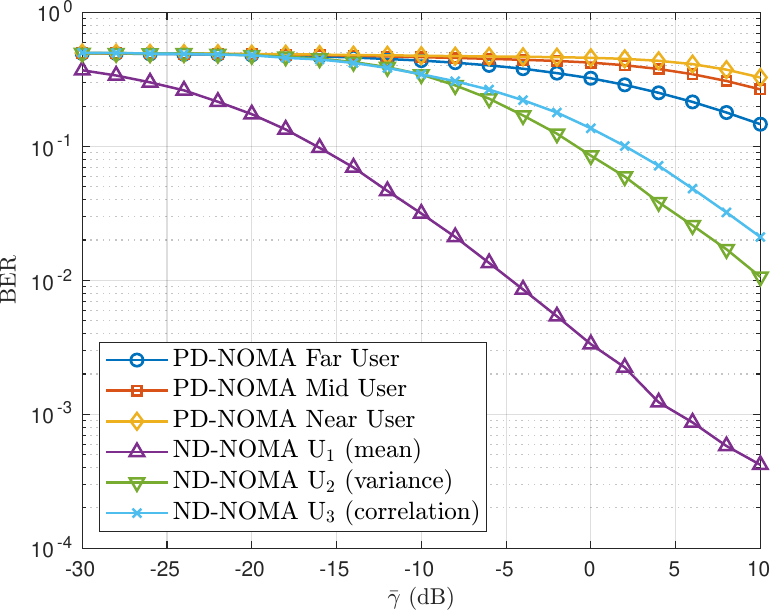}
    \caption{Performance comparison of downlink three-user PD-NOMA and ND-NOMA schemes in terms of BER versus the average SNR $\bar{\gamma}$.}
    \label{fig:fig7}
\end{figure}

Fig. \ref{fig:fig7} compares the BER performance of the proposed three-user downlink ND-NOMA scheme with that of a conventional three-user PD-NOMA system over the same average SNR range, defined as $\bar{\gamma}=P_{\mathrm{tot}}/\sigma_w^2$ with $P_{\mathrm{tot}}=1$ as seen from the \cite{yapici2024noisedomainnonorthogonalmultipleaccess, dos2025off}. This common SNR definition ensures that both schemes are evaluated under identical noise power conditions, enabling a fair comparison.

In the PD-NOMA scheme, users are categorized as near, mid, and far based on their channel conditions, and an asymmetric power allocation is applied, with the far user assigned the highest transmit power. SIC is employed at the receivers of the near and mid users, while the far user performs direct detection. 

In contrast, the proposed ND-NOMA scheme supports three users by embedding information into different statistical dimensions of the transmitted noise signal, namely the mean, variance, and correlation. Since user separation is achieved in the noise domain, ND-NOMA does not require SIC and thus avoids error propagation effects.

The simulation results show that ND-NOMA consistently achieves lower BER than PD-NOMA across the examined $\bar{\gamma}$ range. This performance advantage can be attributed to noise-domain encoding and the absence of SIC, whereas the PD-NOMA performance is strongly influenced by interference cancellation accuracy and power imbalance, particularly at low-to-medium SNR values.

\section{Conclusion}
This work has redesigned the two user ND-NOMA scheme to three users by transmitting information using the mean, variance and also the correlation of the real Gaussian noise. Unlike traditional schemes such as PD-NOMA that rely on SIC, ND-NOMA achieves multi-user access with significantly lower receiver complexity and power consumption. Our analytical and simulation results have shown the reliable performance, especially in Rician fading environments where there are more line of sight components. Mean-based, variance-based and correlation-based systems demonstrate the scalable multi user access property of three user ND-NOMA, which proves that our system is suitable for lower-power, IoT-oriented scenarios. Thanks to its non-complex detection framework, ND-NOMA emerges as a promising solution for future communication networks. As a future direction, investigating the practical feasibility of the three-user ND-NOMA through hardware implementation using software-defined radio (SDR) devices could offer valuable insights into its real-time processing capabilities, energy efficiency, and suitability for embedded and IoT-based applications. Further users could utilize skewness, kurtosis, or higher-order noise statistics.

\appendices
\section{Monte Carlo Integration Method for the Three-User ND-NOMA system}
\label{sec:monte_carlo_threeuser}

Monte Carlo integration is a widely utilized technique for estimating integrals that are difficult or impossible to compute analytically, especially in high-dimensional problems \cite{yapici2024noisedomainnonorthogonalmultipleaccess}. In the context of the three-user ND-NOMA system, the unconditional BEP can be expressed as the following multidimensional integral:

\begin{align}
\bar{P}_b = \int_V g(h_{1,R}, h_{1,I}, h_{2,R}, h_{2,I}, h_{3,R}, h_{3,I}) \, d\hat{h},
\end{align}

where the function \( g(\cdot) \) is defined as:

\begin{equation}
g = P_b \cdot f(h_{1,R}) f(h_{1,I}) f(h_{2,R}) f(h_{2,I}) f(h_{3,R}) f(h_{3,I}),
\end{equation}

and the integration differential is denoted by \( d\hat{h} = d h_{1,R} d h_{1,I} d h_{2,R} d h_{2,I} d h_{3,R} d h_{3,I} \). Here, \( f(\cdot) \) represents the marginal probability density functions of the Rician distributed channel coefficients. The integration region \( V \) spans the entire range of possible channel realizations.

To enhance computational efficiency, we employ importance sampling by introducing a joint sampling distribution that factorizes across all variables:

\begin{equation}
z(\cdot) = \prod_{i=1}^{3} z(h_{i,R}) z(h_{i,I}),
\end{equation}

where \( z(\cdot) \) is selected to closely match the actual distributions of the fading variables.

The integral is then estimated using a Monte Carlo approximation based on \( J \) independently drawn samples \( \mathbf{h}^{(j)} = [h_{1,R}^{(j)}, h_{1,I}^{(j)}, h_{2,R}^{(j)}, h_{2,I}^{(j)}, h_{3,R}^{(j)}, h_{3,I}^{(j)}] \). The estimate becomes:

\begin{equation}
\bar{P}_b \approx \frac{1}{J} \sum_{j=1}^{J} \frac{g(\mathbf{h}^{(j)})}{z(\mathbf{h}^{(j)})}.
\end{equation}

In our simulations, we set \( J = 10^6 \) sample points to ensure high-accuracy estimates for the BEP under various Rician fading conditions in the three-user ND-NOMA system.

\bibliographystyle{IEEEtran}
\bibliography{bib_2023}
\balance

\end{document}